# Drone based superconducting single photon detection system with detection efficiency more than 90%

Ruoyan Ma[1,2], Zhimin Guo[1,2,3], Dai Chen[3], Xiaojun Dai[1,2], You Xiao[1,2], ChengJun Zhang[3], Jiamin Xiong[1,2], Jia Huang[1,2], Xingyu Zhang[1,2], Xiaoyu Liu[1,2], Liangliang Rong[1,2,4], Hao Li[1,2,4], Xiaofu Zhang[1,2,4,5], Lixing You[1,2,4,6]

[1]*National Key Laboratory of Materials for Integrated Circuits, Shanghai Institute of Microsystem and Information Technology, Chinese Academy of Sciences, Shanghai 200050, China*

[2]*Shanghai Key Laboratory of Superconductor Integrated Circuit Technologies, 865 Changning Road, Shanghai 200050, China*

[3]*Photon Technology, 11 Guigu 2nd Road, Jiashan, Zhejiang 314100, China*

[4]*Center of Materials Science and Optoelectronics Engineering, University of Chinese Academy of Sciences, Beijing 100049, China*

[5]*zhangxf@mail.sim.ac.cn*

[6]*lxyou@mail.sim.ac.cn*

**Bounded by the size, weight, and power consumption (SWaP) of conventional superconducting single photon detectors (SSPD), applications of SSPDs were commonly confined in the laboratory. However, booming demands for high efficiency single photon detector incorporated with avionic platforms arise with the development of remote imaging and sensing or long-haul quantum communication without topographical constraints. We herein designed and manufactured the first drone based SSPD system with a SDE as high as 91.8%. This drone based SSPD system is established with high performance NbTiN SSPDs, self-developed miniature liquid helium dewar, and homemade integrated electric setups, which is able to be launched in complex topographical conditions. Such a drone based SSPD system may open the use of SSPDs for applications that demand high-SDE in complex environments.**

## 1. INTRODUCTION

Superconducting nanowire single photon detectors (SSPDs), constituted by meander nanowires with dimensions of sub-100 nm in width and a few nanometers in thickness, have been widely applied in various fields due to their superior detection performance [1-10]. However, owing to the relatively low superconducting critical temperature of conventional superconducting thin films, the high efficiency SSPDs can only be operated at relatively low temperatures, commonly below 2.5 K. Therefore, these SSPD systems are commonly incorporated with Gifford-Mcmahon (GM) cryocoolers or dilution refrigerators. Due to the relatively large size, weight and high-power consumption (SWaP) of the cooling system, the high efficiency SSPDs can hardly be deployed on airborne or spaceborne platforms, let alone a drone based versatile platform [11-15]. Besides the necessary cryocooler, the operation of avionic platforms based SSPDs also necessitates reliable and stable electric bias modules and amplification modules, as well as fast pulse counters with low power dissipations [15]. Due to these requirements, currently, it is still out of reach to launch a SSPD system out of the laboratory.

Recently, there are booming demands for high SSPDs that can be operated with mobile systems [16-20]. For instance, SSPDs have been successfully applied for the light detection and ranging (LiDAR) with extraordinary and temporal resolutions [20-22], due to their high detection efficiency and low timing jitter. Once the SSPDs can be deployed on drone or airborne platforms, it would significantly facilitate the recent airborne single-photon LiDAR systems [23]. Besides, SSPDs are also highly desirable for the deep space optical communication (DSOC) applications [16,17,24-26]. By encoding data in photons at near-infrared wavelengths rather than radio waves, it is expected to increase the data transmission rate by one or two orders of magnitude, completely revolutionize the DSOC technology and facilitating the deep space explorations. In the DSOC demonstration of the Psyche mission, an ultra-high-definition streaming video was sent from Mars with a maximum data downlink rate of 267 Mbps, which is based on 64-pixel SSPD array [27]. Due to the lack of SSPDs that can be equipped on a spacecraft, the uplink rate, however, was limited around 1.6 kbps, leading to an asymmetric space-to-ground communication.

A feasible way to realize these avionic applications is to apply miniature liquid dewar to provide the necessary low temperature operation environment, which is able to get rid of the SWaP limitations. Therefore, high efficiency SSPDs that can be operated at liquid helium temperature become the keystone for drone-based applications [15]. Generally, the operation temperature of SSPDs is an overall effect between the critical temperature $T_c$ and the photoresponse performance of superconducting nanowires. Despite of the higher $T_c$ of superconducting nanowires with lower disorder level or larger film thickness, the single photon detection performances of these nanowires are actually suppressed by the relatively low photon energy transformation efficiency [28-30]. In superconducting nanowires with low disorder, the critical temperature and the critical current can be significantly enhanced, but the photon energy is fastly transferred into the substrate due to the strong electron-phonon interactions, leading to a relatively poor intrinsic detection efficiency (IDE) [30,31]. To simultaneously enhance the IDE and operation temperature of SSPDs, an effective way is to reduce the nanowire width [32]. However, the fabrications of constriction-free narrow nanowires (sub 50 nm) with relatively large sensitive area (with diameter larger than 15 μm) still remains challenging. Another method to improve the IDE at finite operation temperature is to fabricate more sensitive nanowires by increasing the disorder level [33]. Anyway, the strong disorder can significantly suppress the superconducting properties of two-dimensional superconducting films, which may even drive the superconductor into insulators at the quantum critical disorder level [34]. Consequently, the key to improve the operation temperature and detection efficiency of SSPDs is to balance the disorder level and superconducting properties of superconducting thin films.

In this letter, we successfully demonstrated the fabrications of high efficiency SSPDs based on highly disordered NbTiN films. The NbTiN-SSPD shows a SDE more than 90% operated at liquid helium temperature. To get rid of the conventional cryocoolers and build a lightweight detection system, we also designed and manufactured a miniature liquid helium dewar, which is compatible with commercial drones and other airborne platforms. Incorporating with home-made electric setup for biasing and data storage, we successfully constructed a high efficiency drone-based SSPD system, which is applicable for remote imaging and sensing, or long-distance quantum communications.

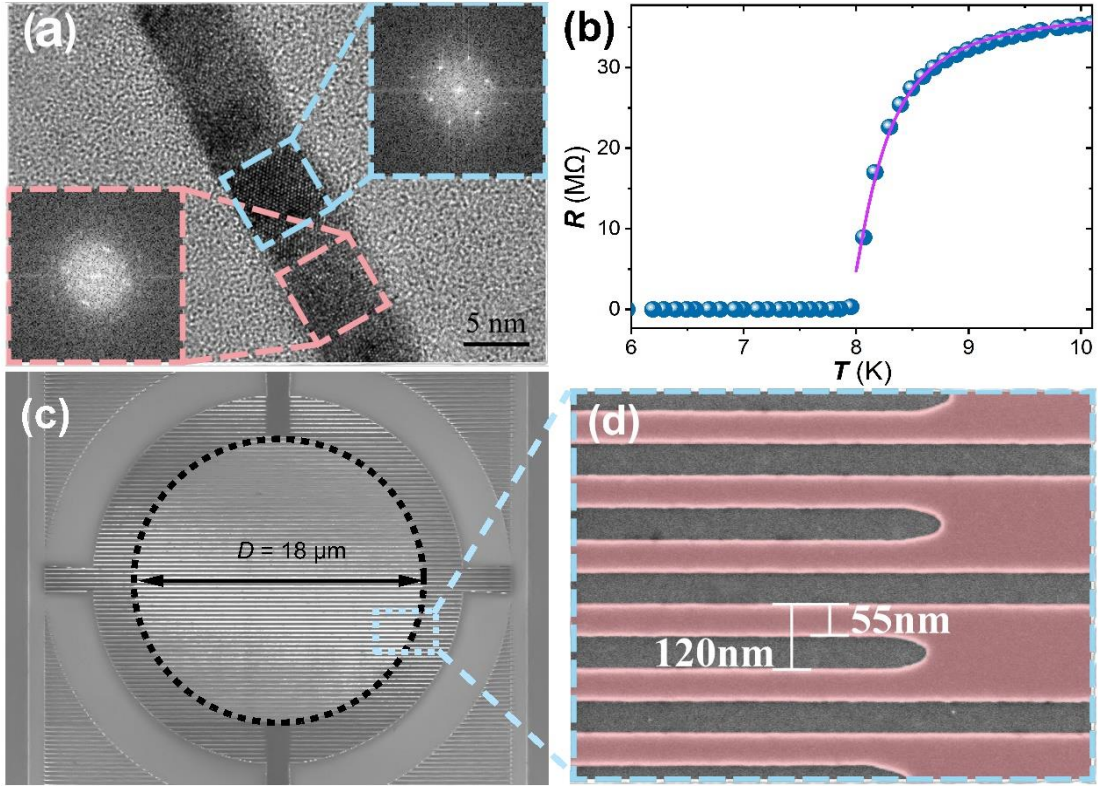

**Fig. 1 (a)** The transmission electron microscope image of the involved NbTiN superconducting thin film. **(b)** The superconducting to normal transitions for a 55-nm-wide detector. **(c)** The scanning electron microscope image of the fabricated SSPDs, where the diameter of the sensitive area is 18 μm. **(d)** The uniformity of the nanowire width.

## 2. DEVICE DESIGN AND FABRICATION

Based on the recently reported disorder-tuning method [33,35], we further optimized the deposition conditions of NbTiN films to realize more sensitive superconducting nanowires. Figure 1(a) shows a transmission electron microscope (TEM) image for the 7-nm-thick granular NbTiN films, in which the average granule size of the resulting NbTiN films is around 5 nm. Moreover, some of the NbTiN granules are even close to the non-crystalline states (Inset of Fig. 1(a)). After the deposition of highly disordered NbTiN films on an alternative $SiO_2/Ta_2O_5$ dielectric mirror (optimized for the telecom wavelength photons), the narrow nanowires are defined by a 100 keV electron beam lithography system by using the positive ZEP520 e-beam resist, and the pattern was then transferred by dry etching using reactive ions of $CF_4$ plasma, as shown in Fig. 1(c) and (d). To guarantee a sufficient coupling efficiency of the devices, the sensitive area of the detector is designed to be 18 μm in diameter (which is two times larger than the beam waist of the fiber). To assure the sensitivity and yields of constriction-free detectors, the nanowire width and pitch are still respectively designed to be 55 nm and

120 nm, leading to a filling factor around 46%. The inner corners of meander turns are elliptically optimized to reduce the dark count rate [36,37], as it is shown in Fig. 1(d). Figure 1(b) presents the temperature dependence of resistance $R(T)$ for the fabricated devices. The $R(T)$ dependence of the NbTiN devices can be well-described with the one-dimensional superconducting fluctuation mode [38], and the zero-resistivity critical temperature is around 8 K.

To characterize the detection performance of the fabricated SSPDs at liquid helium temperature, we here firstly measured the SDE as a function of temperature in a Gifford-McMahon (GM) cryocooler. Figure 2 presents the measured bias current dependence of SDE at temperatures of 2.2 K, 3 K, 3.5 K, and 4.2 K, respectively. Due to the increased disorder level, the critical current at 2.2 K is slightly lower than the previous reported devices with the same device configuration [15,33]. At 2.2 K, the detector shows a saturated SDE, with a maximum SDE more than 95% for 1550 nm photons. With the increasing temperature, due to the gradually decreasing switching current, the maximum SDE is suppressed to 91.8% at 4.2 K. Compared with the previously reported SSPDs operated at liquid helium temperature [15,33,39,40], although the critical current is suppressed, the SDE more than 90% demonstrates that the SSPD system can be applied for building drone-based or avionic high efficiency single photon detection platform without significantly degrading its sensitivity.

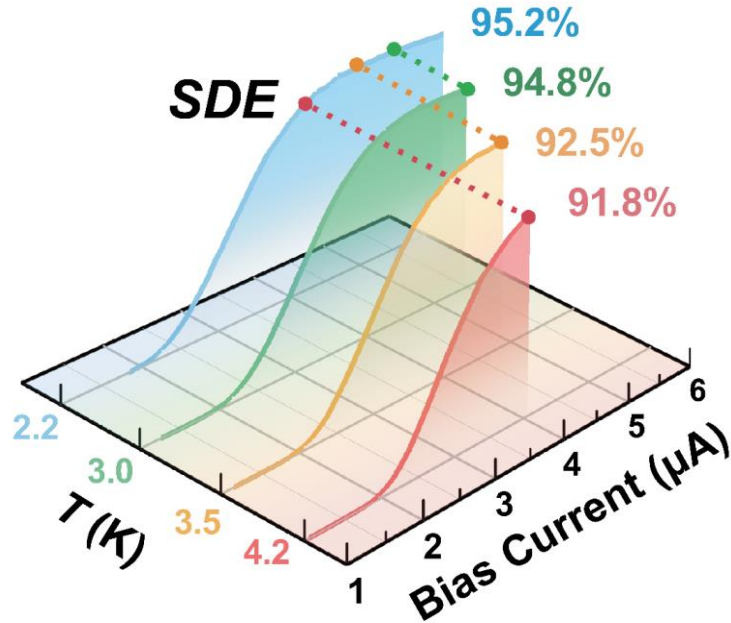

**Fig. 2** The temperature dependence of the system detection efficiency as a function of bias current.

## 3. DRONE-BASED SYSTEM DESIGN AND MANUFACTURE

To realize a drone-based superconducting single photon detection system that can be launched readily, a commercial drone with finite size and power consumption was adopted as the airborne platform. Due to the finite space for mounting the whole SSPD system, the size, weight, and power consumption (SWaP) of the cryostat that supporting the necessitated low temperature working environment for SSPDs must be low enough.

To this end, a miniaturized liquid helium Dewar was designed and manufactured based on our previously developed mobile SSPD system [15]. For easy attachment to the drone, the Dewar's height and diameter were specifically designed to be 450 mm and 250 mm respectively, with a maximum helium capacity of 3 liters, and an overall weight of approximately 12 kg. To ensure a relatively long operational time, an intercalation shielding layer made of thin aluminum is connected to the neck of the inner liquid helium vessel and is positioned in the vacuum space between the inner and outer vessels, despite of the limited space. Additionally, 45 layers of multilayer insulation were wrapped around the inner vessel wall to further enhance insulation. To minimize conduction heat loss along the neck of the inner vessel, a specially designed bellows tube was welded between the bottom and top of the neck. The combined weight of the inner vessel and the intercalation shielding layer is less than 2 kg.

For the installation of SSPDs, we designed a dipstick made from a glass fiber reinforced plastic tube with baffles and polyurethane foam, which is able to install two SSPDs simultaneously. It is noteworthy that the baffles and polyurethane foam were strategically positioned at a height that aligns with the neck of the inner vessel. This arrangement helps to effectively distribute the helium vapor and minimize radiation and convection heat loss. To reduce the heat load from the room temperature end and maintain constant pressure inside the inner vessel, a leakage capillary tube for the evaporated helium gas was welded onto the dipstick.

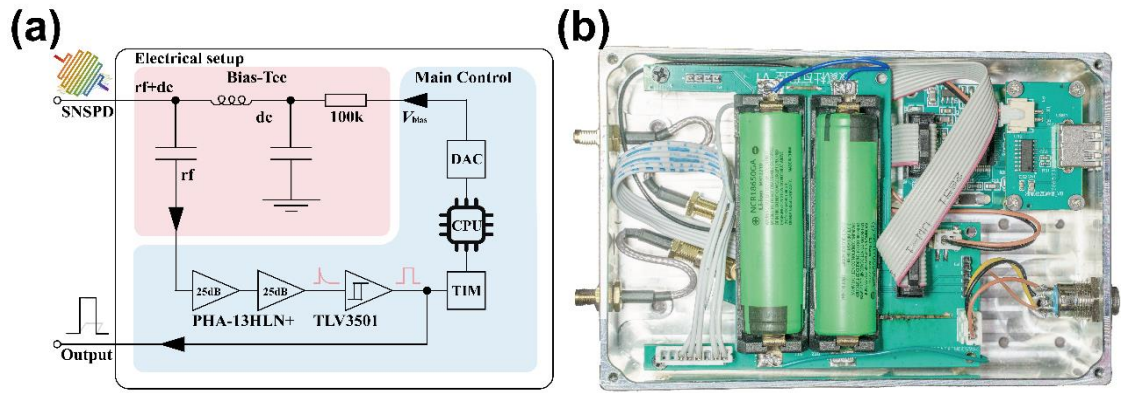

**Fig. 3 (a)** A sketch for the electrical setup integrating the bias circuit, the amplification circuit, and pulse-converter and counter. **(b)** The manufactured electrical setup.

With respect to the bias and readout circuit of the SSPD system, we here designed and manufactured an integrated electrical setup, which combines the bias circuit, amplification module, pulse-counting module, and the data storage module. Figure 3 shows manufactured electric setup, which is power supplied by two NCR18650A lithium-ion batteries. The bias current on the SSPD is generated by the $V_{bias}$ through an adjustable resistor (100 kΩ), where the programable $V_{bias}$ is modulated by the main control unit (MCU) and a digital-to-analog converter. The detection signal from the SSPDs is amplified by two low-power amplifiers, with a total gain more than 50 dB. Then, these pulse voltage signals are converted into transistor-transistor logic signal by a high-speed comparator, which either can be outputted to a standard counter or be directly counted by the Timer module of the MCU. Finally, the bias current dependence of the SDE, or the counts at a fixed bias current, are saved in a text file. Figure 3(b) shows the manufactured electric setup, which is 150 mm in length and 100 mm in width. With fully charged lithium-ion batteries, the electric setup is able to sustain for more than 30 hours, with a total weight around 1 kg.

## 4. DRONE-BASED SSPD SYSTEM AND RESULTS

Based on the manufactured miniature liquid helium dewar, we here constructed the drone-based SSPD system based on the T40 drone (AGRAS T40, DJI agricultural drone), as it is shown in Fig. 4(a). Firstly, two aluminum arms are fixed on the drone, and the dewar is then mechanically mounted on the two arms. The electrical setup for testing and data-recording is also fixed on the aluminum arm. After the establishment of the whole system, we then compared the measured bias current dependence of SDE on the ground for the drone-based SSPD system with the measured SDE from standard measurement setups, as it is shown in Fig. 4(b). The SDE in the liquid helium dewar is well consistent with that measured in the GM cryocooler, demonstrating the robust detection performance of the NbTiN SSPDs. Moreover, the completely consistent optical and electric response also demonstrate the reliable electrical supports of the home-made SSPD electrical measurement setup.

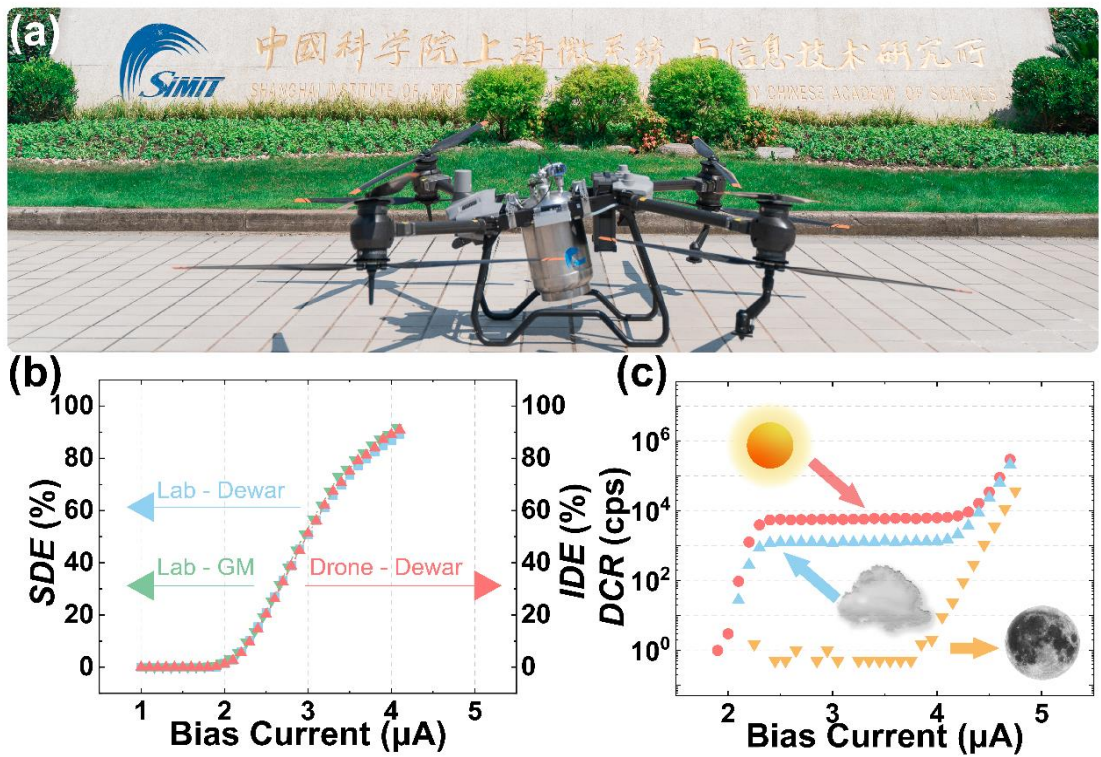

**Fig. 4 (a)** The drone-based superconducting nanowire single photon detection system. **(b)** The bias current dependence of the SDE measured in a GM cryocooler (light green), and in the miniature liquid helium dewar (light blue). The measured IDE as a function of bias current of the drone-based SSPD system, with a height of 30 m. **(c)** The dark count rate as a function of bias current under different operation environment.

To characterize the detection performance on the drone, we here measured the intrinsic detection performance of the system since currently a receiver is absent on the drone. The SSPDs are illuminated with a semiconductor laser diode with a fixed attenuation factor, where the laser source is also mounted on the drone. Before launching the drone, the attenuation factor was set to assure that the count rate (CR) exceeded 1.5 million counts per second, which is roughly three orders of magnitude more than the dark count rate (DCR), ensuring the accuracy of the IDE, which is defined by $IDE = (CR - DCR)/CR$. The drone based SSPD system was then launched and hovered at an altitude of 30 m from the ground (which is limited by the local statutory regulations in Shanghai), and the IDE was simultaneously measured under both flight conditions and hovering conditions. The measured IDE as a function of bias current is presented in Fig. 4(b), which is also coincide well with the measured data on the ground, further demonstrating the reliability and stability of the NbTiN detector and the drone-based SSPD system.

The most significant challenge for the operation of the drone-based SSPD system is the relatively high dark counts. Figure 4(c) compares the bias current dependence of the DCR under different operation conditions. In a dark operation environment, the DCR of the drone-based SSPD system is consistent with that operated in a standard GM cryocooler, where the DCR was measured by blocking the fiber and turning off the laser source. The DCR is around 1 cps when the bias current is relatively low, and increase exponentially in the intrinsic regime. However, under a bright sunlight environment with a ground temperature more than 50 degree centigrade, the DCR was increased to around ten thousand cps, which is due to the strong stray photons and the black-body radiation from the high temperature terminal of the optical fiber. It is also interesting to note that when hovering the drone in a shadow environment (where the bright sunlight is shaded occasionally by the cloud), the DCR was then effectively reduced by an order of magnitude. As a consequence, by optimizing the design of the receiving and coupling components of the drone-based SSPD system, the system dark count rate can be significantly lowered, even operating the drone-based SSPD system at a relatively high altitude with strong sunlight.

Finally, to further improve the detection performance of a drone-based SSPD system, one possible way is to continuously optimize the nanowire geometrics, where the superconducting properties can be optionally tunned. By slightly varying the film thickness, nanowire width, and the filling factor of SSPDs, the detection efficiency and the optimal operation temperature can be further enhanced. With respect to the relatively high dark counts, a narrow-band pass filter can be applied to filter out the stray photons and the blackbody radiations outside the targeted wavelength [41]. Such a narrow-band pass filter can be deposited either directly on the SSPD chips or on the end-face of the fiber. Concerning the miniature liquid helium dewar, currently most of the weight is still concentrated on the steel outer vessel, which provides the necessary strength against harsh operation environments. Anyway, the weight can be further reduced by applying spaceflight aluminum-titanium alloy.

## 5. CONCLUSION

In summary, we successfully constructed a commercial drone-based superconducting nanowire single photon detection system. By optimizing the deposition condition and nano-fabrication process, we have fabricated NbTiN-SSPDs with near-saturated system detection efficiency. At liquid helium temperature, the system detection efficiency is as high as 91.8%. Furthermore, a specially designed miniature liquid dewar is manufactured, which can be directly incorporated in the T40 drone. To resolve the readout and data-storage of SSPDs, we also designed and manufactured special electrical setups. The whole system except the drone is merely power-supplied by two lithium batteries. Such a high efficiency drone-based SSPD system is promising for applications of three-dimensional high-resolution imaging and sensing, or avionic quantum communications. Moreover, such lightweight and low power systems are even able to be installed on spaceborne platforms, further facilitating the deep space optical communications and space explorations.

**Funding.** National Key Research and Development Program of China (No. 2023ZD0300100, 2023YFB3809600, & 2023YFC3007801) and the National Natural Science Foundation of China (No. 62301543).
**Acknowledgment.** The fabrication was performed in the Superconducting Electronics Facility (SELF) of SIMIT, CAS. We thank Danhua Wu for technical assistance and Lu Zhang for assisting in thin film depositions.
**Disclosures.** The authors declare no conflicts of interest.
**Data availability.** Data underlying the results presented in this paper may be obtained from the authors upon reasonable request.